\documentclass[prl,twocolumn,aps,showpacs]{revtex4}
\usepackage{graphicx}
\usepackage{amsmath}
\usepackage{amsfonts, amssymb, amstext}

\def\q{{\bf q}}

\def\x{{\bf x}}

\newcommand{\vecb}[1]{{\bf #1}}

\newcommand{\av}[1]{\langle #1 \rangle}
\def\ltsim{\vbox {\hbox{\lower .8\baselineskip \hbox{$<$}} \break
                 \hbox{\lower 0.2\baselineskip \hbox{$\sim$}} } }

\begin{document}

\title{Dissipative Quantum Ising model in a cold atomic spin-boson mixture}

\author{Peter P. Orth, Ivan Stanic, and Karyn Le Hur}
\affiliation{Department of Physics, Yale University, New Haven, CT 06520, USA}
 
 \date{\today} 

\begin{abstract}
Using cold bosonic atoms with two (hyperfine) ground states, we introduce a spin-boson mixture which allows to implement the quantum Ising model in a tunable dissipative environment. The first specie lies in a deep optical lattice with tightly confining wells and forms a spin array; spin-up/down  corresponds to occupation by one/no atom at each site. The second specie forms a superfluid reservoir. Different species are coupled coherently via laser transitions and collisions. Whereas the laser coupling mimics a transverse field for the spins, the coupling to the reservoir sound modes induces a ferromagnetic (Ising) coupling as well as dissipation. This gives rise to an  order-disorder quantum phase transition where the effect of dissipation can be studied in a controllable manner.
\end{abstract}

\pacs{03.75.Mn, 64.70.Tg, 71.27.+a}
\maketitle

%\section{Introduction}
Spin-boson models are essential ingredients in 
quantum optics \cite{Milburn}, nuclear physics \cite{Klein}, quantum chaos \cite{Haake}, and quantum dissipation \cite{leggett,Weiss}. In particular, an ensemble
of identical two-level systems, each coupled to common radiation field modes, the Dicke model \cite{Dicke}, was introduced initially to describe the superradiant emission (a sudden increase in the rate of coherent spontaneous emission of an ensemble of atoms). This model exhibits an interesting
order-disorder transition for the two-level systems both in the limit of only one or several boson modes \cite{Tolkunov}. In this Letter, we theoretically envision a different ``spin-boson'' mixture, {\it i.e.}, a spin array coupled to a large collection of harmonic oscillators, realized using cold atomic bosons, which allows us to explore how the properties of the celebrated quantum Ising model and of the emergent quantum phase transition for the spins \cite{Subir} are modified in a tunable bosonic environment. We emphasize that our setup embodies the first tunable realization of the quantum Ising model in a dissipative bath, and that several critical exponents can be measured using standard imaging techniques. Generally, mixtures of different species of cold atoms open a fascinating field of many-body physics where exotic spin systems and quantum phase transitions can be engineered and probed 
\cite{Zoller1,Demler,Jaksch}. %It should be noted that a two-component bosonic mixture has already been realized experimentally \cite{Catani}.

%Dissipation and its interplay with a (non-local) Ising interaction, mediated by the bath, in this well controlled and clean mixture of optically trapped cold bosonic atoms is of wide interest not only because of the emergence of an unusual (dissipative) quantum phase transition but also for its implications to entanglement and quantum computation\cite{Zoller1,Demler,Jaksch}. 

%
\begin{figure}[t]
\begin{center}
\includegraphics[width=\linewidth]{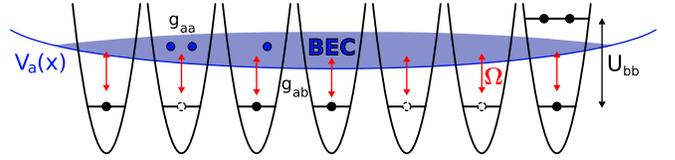}
\end{center}
\caption{Setup comprising a BEC and two-level systems.}
\label{fig:1}
\end{figure}

More precisely, we consider cold bosonic atoms with two (hyperfine) ground states $a$ and $b$, trapped by different state-selective external potentials~\cite{Jaksch2}. Whereas atoms in state $a$ form a Bose-Einstein Condensate (BEC) of dimension $d=(1,2,3)$, held in a shallow potential $V_a(\x)$, the $b$ atoms are trapped in a deep $d$-dimensional optical lattice with well-separated lattice sites, so that hopping between different sites can be neglected (see Fig.~\ref{fig:1}). We consider the collisional blockade regime of large onsite interaction $U_{bb}$, where only states of occupation number $n_b=0,1$ per lattice site contribute to the dynamics. With all higher occupied states being adiabatically eliminated, one obtains a $d$-dimensional array of two-level systems or quantum $1/2$-pseudospins. Spin-up/down corresponds to an occupation number of one/zero. The spins are coupled to the low-energy phonon excitations (superfluid sound modes) of the BEC in two ways: via collisions between $a$ and $b$ atoms and via laser mediated Raman transitions between the two hyperfine species. The dynamics of a single two-level system embedded in a BEC has been addressed in Ref. \cite{Zoller2} and the emergent quantum phase transition can be understood in terms of the spin-boson model \cite{leggett,Weiss}, which has abundant practical applications in physics \cite{Schon, Karyn}. 

The collisional interaction of the atoms is described by a contact pseudopotential with coupling strength $g_{\alpha\beta}=4 \pi a_{\alpha\beta} \hbar^2/m$ for $\alpha,\beta\in \{a,b\}$, where $a_{\alpha\beta}$ is the corresponding s-wave scattering length and $m$ the atomic mass. As shown below, the 
cross-coupling $g_{ab}$ between the spin array of $b$ atoms and the superfluid reservoir of $a$ atoms induces a ferromagnetic (short-range) Ising exchange interaction between spins. On the other hand, at each site, the Raman coupling acts as a transverse field $\Delta$, favoring paramagnetism. In addition, the coupling of the spins to the long-wavelength BEC-phonons introduces dissipation to the dynamics of the spin system. Integrating out the BEC (phonon) degrees of freedom, we demonstrate that this will result in a dissipative quantum Ising model (DQIM) for the spins \cite{Matthias,Matthias2,Georges}, that exhibits an unusual dissipative order-disorder phase transition. One can determine the phase of the spin system (ferro- or paramagnetic) by measuring the mean occupation number $\av{n_b}$ of the lattice sites \cite{Grangier}. In addition, measuring $\av{n_b}$ for different Raman laser intensities and detunings allows to extract several critical exponents.

%\section{Model}
{\it Model.--} The $a$ atoms form a BEC (which can have dimension $d=1,2,3$), and the
Bose-field operator can be split into magnitude and phase $\hat{\psi}_a(\vecb{x}) = \sqrt{\hat{\rho}_a(\vecb{x})} e^{- i \hat{\phi}(\vecb{x})}$. We consider the low temperature limit where the superfluid density is equal to the equilibrium atomic density $\rho_a$. Introducing the density fluctuation operator $\hat{\Pi}(\vecb{x}) = \hat{\rho}_a(\vecb{x}) - \rho_a$, which is canonically conjugate to the superfluid phase $\hat{\phi}(\vecb{x})$, the dynamics of the BEC in the long-wavelength approximation is then described through a quantum-hydrodynamic Hamiltonian,
\begin{equation}
  \label{eq:1}
  H_a=\int d\vecb{x} \left( \frac{\hbar^2\rho_a}{2m} | \nabla \hat{\phi}(\vecb{x})|^2 + \frac{m v_s^2}{2 \rho_a} \hat{\Pi}^2(\vecb{x}) \right)\,, 
 \end{equation}
 where $v_s$ is the sound velocity and $m$ the atomic mass. This description can be applied to all dimensions $d=(1,2,3)$. In $d=(2,3)$, one obtains this Hamiltonian resorting to the Gross-Pitaevskii approach where the interaction coefficient $g_{aa}$ enters the Hamiltonian via the sound velocity $v_s^2 = \rho_a g_{aa}/m$. In the one-dimensional (1D) case of a tight transverse confinement with frequency $\hbar \omega_\perp \gg (k_B T, \mu$) (with $\mu$ being the chemical potential), which is particularly interesting as it leads to ohmic dissipation, the proper derivation follows the 
 Haldane-Luttinger approach~\cite{Haldane}. In fact, one may identify $K=\hbar \pi \bar{\rho}_a/(m v_s)$ as the Luttinger parameter, where $\bar{\rho}_a = \rho_a l_\perp^2$ is the 1D density that depends on the transverse confinement length $l_\perp = (\hbar/m \omega_\perp)^{1/2}$. The corresponding 1D interaction coefficients are related to the 3D scattering lengths by $\bar{g}_{\alpha\beta} = 2 \hbar \omega_\perp a_{\alpha\beta}$ if $a_{\alpha\beta} \ll l_\perp$~\cite{Olshanii}. 
 
The Hamiltonian $H_a$ can be diagonalized using phonon operators via $\hat{\phi}(\vecb{x}) = i \sum_{\vecb{q}} |\frac{m v_s}{2\hbar {\bf q} V \rho_a}|^{1/2} e^{i {\bf q}{\bf x}}(\hat{\gamma}_{\bf q}-\hat{\gamma}_{-{\bf q}}^{\dagger})$ and $\hat{\Pi}({\bf x}) = \sum_{\bf q} |\frac{\hbar \rho_a {\bf q}}{2 v_s V m}|^{1/2} e^{i {\bf q}{\bf x}} (\hat{\gamma}_{\bf q} +\hat{\gamma}_{-{\bf q}}^{\dagger})$, 
%$\hat{b}_{\q}$, $\hat{b}^{\dagger}_\q$ 
%\begin{equation}
%  \label{eq:2}
%  \begin{split}
%  \hat{\phi}(\vecb{x}) &= i \sum_{\vecb{q}} \left|\frac{m v_s}{2\hbar {\bf q} V \rho_a} \right|^{1/2} e^{i {\bf q}{\bf x}} \left(\hat{b}_{\bf q}-\hat{b}_{-{\bf q}}^{\dagger}\right) \\
%  \hat{\Pi}({\bf x}) &= \sum_{\bf q} \left|\frac{\hbar \rho_a {\bf q}}{2 v_s V m}\right|^{1/2} e^{i {\bf q}{\bf x}} \left(\hat{b}_{\bf q} +\hat{b}_{-{\bf q}}^{\dagger}\right)\,,
%  \end{split}
%\end{equation}
where $V=L^d$ is the volume of the (BEC) system. Since the low-energy excitations of a BEC are phononlike only for wavelengths larger than the healing length $\xi=\hbar/\sqrt{2 m \rho_a g_{aa}}$, the sum is cut off at a frequency $\omega_c = v_s/\xi = \sqrt{2} g_{aa} \rho_a/\hbar$. One gets the well-known commutation relations,
\begin{equation}
\label{eq:2}
[\hat{\phi}(\x), \hat{\Pi}(\x')] = \frac{i}{V}\sum_{\bf q} e^{i{\bf q}({\bf x}-{\bf x}')}= i \frac{\eta(|\x-\x'|/\xi)}{\pi \xi}, 
\end{equation}
where each spatial component of the wavevector runs between $L^{-1} \leq |q_i| \leq \xi^{-1}$, and $\eta(|\x-\x'|/\xi)$ is a function which decays very rapidly for separations larger than the healing length. For instance in $d=1$ one finds, $\eta(x_{jk}) \sim \sin(x_{jk})/x_{jk}$ with $x_{jk} = |{\bf x}_j - {\bf x}_k|/\xi$. Note that for $\xi\rightarrow 0$, one checks that $[\hat{\phi}(\x), \hat{\Pi}(\x')]\rightarrow i\delta(\x -\x')$).

The Hamiltonian $(H_b + H_{ab})$ that describes the dynamics of the $b$ atoms as well as the coupling between the two species (collisions and the Raman coupling), reads \cite{Zoller2}:
\begin{equation}
\begin{split}
  &H_b + H_{ab} = \sum_i \left[ -\hbar r_0 + \int d\x |\psi_{b,i}(\x)|^2 g_{ab} \hat{\rho}_a(\x) \right]  \hat{b}_i^{\dagger} \hat{b}_i \\&+ \int d\x \hbar \Omega \left[\hat{\psi}_a(\x) \psi_{b,i}(\x) \hat{b}_i^{\dagger} + \text{h.c.}\right] + U_{bb} \hat{b}^{\dagger}_i \hat{b}^{\dagger}_i \hat{b}_i \hat{b}_i\,.
\end{split}
\end{equation}
Here, $\hat{b}^{\dagger}_i$ creates a particle in the localized Wannier ground state $\psi_{b,i}$ at lattice site $i$ which has a spatial extent $l_b$. The operator $\hat{\psi}_a(\x)$ annihilates an atom (in state $a$) of the superfluid reservoir at position $\x$. Transitions between the two hyperfine states are induced by Raman lasers with Rabi frequency $\Omega$ and detuning $r_0$. Atoms are transferred from the reservoir to the lowest vibrational band of the lattice and vice versa. In coupling the $b$ atoms to a coherent matter-wave field $\hat{\psi}_a(\x)$, we have assumed that the number of $a$ atoms inside a lattice well is large, $n_a = \rho_a l_b^3 \gg 1$. We also include the collisional interaction between $a$ and $b$ atoms. The repulsive onsite interaction $U_{bb}\approx g_{bb}/l_b^3>0$ of $b$ atoms can be varied using a magnetic or optical Feshbach resonance~\cite{Zoller2}. We assume $U_{bb}/\hbar \gg (\Omega, r_0)$ in order to neglect doubly and higher occupied states (collisional blockade limit).    

Using Pauli matrix notations to describe the spin array, we replace the occupation number operator $\hat{b}_i^{\dagger} \hat{b}_i$ by $(1 + \sigma_i^z)/2$ and $\hat{b}_i^{\dagger} \rightarrow \sigma_i^+$. Furthermore, in the spin-phonon coupling term, we can replace the phonon field operators by their values at the center of the lattice sites $\x_i$, because the phonon wavelength is much larger than the extent of the Wannier ground state $l_b$. 
Neglecting constant terms we obtain the Hamiltonian,
\begin{eqnarray}
  \label{eq:3}
H_b + H_{ab}  &=& \sum_i \left(-\frac{\hbar r}{2} +\frac{g_{ab}}{2}\hat{\Pi}({\bf x}_i) \right)\sigma_i^z \\ \nonumber
&+&\frac{\hbar\Delta}{2}\left(1 + \frac{\hat{\Pi}({\bf x}_i)}{2\rho_a}\right)\left(\sigma_i^+ e^{-i\hat{\phi}({\bf x}_i)}+\text{h.c.}\right)\,,
\end{eqnarray}
where the sum runs over all lattice sites and $\Delta \sim n_a^{1/2} \Omega$ is an effective Rabi frequency which grows with the number of $a$ atoms inside one lattice well. This form of the Rabi coupling requires that the cutoff frequency $\hbar \omega_c  \approx \rho_a g_{aa} \gg \hbar \Delta$, since we only consider ``single-particle'' hopping events. We have also introduced the effective detuning $\hbar r =\hbar r_0 - \rho_a g_{ab} + (\hbar \Delta)^2/(2 U_{bb})$ which includes a collisional mean field shift as well as a shift due to the virtual admixture of doubly occupied states. 

{\it Phonon integration for the mixture.-}To find the dynamics of the spin array, we shall integrate out the phonon modes along the lines of a single two-level system embedded in a BEC~\cite{Zoller2}. In the case of a spin array coupled to bosons in the superfluid phase, we show that the integration of phonons can be still carried out exactly, which is in contrast to the case of a spin array coupled to conduction electrons (Kondo lattice)~\cite{Georges}.

Firstly, we apply the unitary transformation $H'=U^{\dagger}HU$ with $U=\exp\left[-i\sum_j \sigma_j^z \hat{\phi}({\bf x}_j)/2\right]$ to the total Hamiltonian $H= (H_a + H_{ab} + H_b)$ in order to eliminate the operator $\exp(-i\hat{\phi}({\bf x}_i))$ from the Raman term. As a result, the spin-bath coupling takes the exact form,
\begin{equation}
  \label{eq:2a}
  \sum_j  \left[ \left( -1 + \frac{g_{ab}}{g_{aa}} \right) \sigma_j^z - \frac{\hbar \Delta}{2 \rho_a g_{aa}} \sigma_j^x \right]\frac{g_{aa} \hat{\Pi}(\x_j)}{2 }\,,
\end{equation}
where the $\sigma^x$-term can be neglected as long as one stays away from the point $g_{ab} = g_{aa}$ since $\Delta \ll \omega_c$. As we show below, we can only reach the interesting regime of the quantum phase transition if $|g_{ab}| \ll g_{aa}$, since the ferromagnetic interaction dominates otherwise. Thus, we assume that $g_{ab}$ is sufficiently small and keep only the coupling term of the $z$-component of the spin to the bath. 

The unitary transformation also produces a spin-spin exchange term, and the total Hamiltonian becomes,
\begin{equation}
  \label{eq:6}
  \begin{split}
    H' & = - \sum_j \frac{\hbar \Delta}{2} \sigma_j^x + \sum_{j,\q} \left[ - r + \lambda_{\bf q} e^{i{\bf q}{\bf x}_j}\left(\hat{\gamma}_{-{\bf q}}^{\dagger} +\hat{\gamma}_{{\bf q}} \right) \right]  \frac{\hbar \sigma_j^z}{2}  \\ & +\sum_{j,k} \frac{ \rho_a g_{aa} \eta_{jk}}{8 \pi \xi \rho_a} \left[\left( 1 - \frac{2 g_{ab}}{g_{aa}} \right) \sigma_j^z \sigma_{k}^z  \right] + \sum_{\bf q} \hbar\omega_{\bf q} \hat{\gamma}^{\dagger}_{\bf q} \hat{\gamma}_{\bf q}\,,
  \end{split}
\end{equation}
where we have written $\eta(|\x_j - \x_k|/\xi) = \eta_{jk}$ and $\omega_{\bf q}=v_s|{\bf q}|$. For realistic conditions, the separation between neighboring lattice sites $d_b$ is of comparable size as the healing length ($\xi\approx d_b>l_b$) \cite{Zoller2}, and since the function $\eta_{jk}$ defined in Eq.~(\ref{eq:2}) decreases very rapidly for separations larger than $\xi$, below we can safely restrict ourselves to nearest-neighbor spin interactions. We have defined:
\begin{equation}
\lambda_\q = \left(\frac{g_{ab}\rho_a}{m v_s^2}-1\right)\left|\frac{m v_s^3 {\bf q}}{2 V\rho_a \hbar}\right|^{1/2}.
\end{equation}
Now, the integration of phonon modes can be carried out exactly (using coherent state functional integrals). This produces an extra term in the action of the spin array,
\begin{multline}
  \delta S' = - \frac{\lambda_\q^2}{4 \beta \omega_\q} \sum_{\q; n=-\infty}^{\infty} \int_0^{\hbar \beta} d\tau d\tau' \frac{\omega_\q^2 - i \omega_\q \Omega_n}{\omega_\q^2 + \Omega_n^2} \\ \times e^{i \Omega_n (\tau - \tau')} \sigma_{\bf q}(\tau) \sigma_{\bf q}^*(\tau')\,,
\end{multline}
where $\beta=(k_B T)^{-1}$, $\sigma_{\bf q}(\tau) =\sum_j \sigma_j^z(\tau) e^{i \q \x_j}$, $\Omega_n=2\pi n/(\hbar\beta)$ are Matsubara frequencies, and $\lambda_\q^2/\omega_\q$ does not depend on $\q$.  
In general, the coupling to the boson bath provides two distinct contributions, which can be identified through, $\sigma_{\bf q}(\tau) \sigma_{\bf q}^*(\tau') = \frac12 (\sigma_{\bf q}(\tau) \sigma_{\bf q}^*(\tau) + \sigma_{\bf q}(\tau') \sigma_{\bf q}^*(\tau')) - \frac12 (\sigma_{\bf q}(\tau) - \sigma_{\bf q}(\tau')) (\sigma_{\bf q}^*(\tau) - \sigma_{\bf q}^*(\tau'))$ \cite{Weiss}. 

The term which is local in time simply renormalizes the spin-spin Ising interaction.  Adding up all spin interaction terms, we find that the overall interaction is ferromagnetic and is limited to nearest neighbors (as the healing length $\xi \approx d_b$): $  \sum_{\av{j,k}} \left[J \int d\tau \;  \sigma_j^z(\tau)  \sigma_{k}^z(\tau)\right] $ with $J = -g_{ab}^2/(8 \pi g_{aa} \xi)<0$. In contrast to itinerant electrons which produce an RKKY interaction between spins which can be either ferromagnetic or antiferromagnetic, dependently on the electron density, here phonons result in a ferromagnetic Ising exchange \cite{Privman}. We have checked this result by setting $\Delta=0$ and by applying a different unitary transformation $U'=\exp[-i \sum_j \frac{g_{ab}}{2 g_{aa}}\sigma_j^z \hat{\phi}(\x_j)]$ to the Hamiltonian $H$ in accordance with Refs. \cite{Cirac,Privman}.  Similar to the case of a single two-level system coupled to a bosonic bath \cite{Weiss,note},  the second (dissipative) contribution stems from the pole at 
$\omega_{\bf q}=-i\Omega_n$ (or $\omega_{\bf q}=i\Omega_n$). Finally, the spin dynamics is dictated by the action \cite{note},
\begin{equation}
  \label{eq:31}
  \begin{split}
    S_{\text{spin}} &=  - \sum_j \int_0^{\hbar \beta} d\tau \left[ \frac{\hbar \Delta}{2} \sigma_j^x(\tau) + \frac{\hbar r}{2} \sigma_j^z(\tau) \right] \\
    & - \frac{1}{8} \sum_j \int_0^{\hbar \beta}  d\tau d\tau' \alpha(\tau - \tau') \sigma_j^z(\tau) \sigma_j^z(\tau')\\
&+J  \sum_{\av{j,k}} \int_0^{\hbar \beta} d\tau \;  \sigma_j^z(\tau)  \sigma_{k}^z(\tau);
 \end{split}
\end{equation}
$\alpha(\tau - \tau') = \frac{1}{\pi} \int_0^\infty d\omega J(\omega) e^{- \omega |\tau - \tau'|}$ involves the spectral density $J(\omega) = \hbar \pi \sum_\q |\lambda_\q|^2 \delta(\omega - \omega_\q) = 2 \pi \hbar \alpha \omega^d$ $(i\Omega_n\rightarrow \omega+i0^+)$ and the dimensionless dissipative parameter
\begin{equation}
  \label{eq:31}
  \alpha = \frac{D_0(d) m v_s^2}{4 \hbar (2 \pi)^d v_s^d \rho_a} \left(\frac{g_{ab}\rho_a}{m v_s^2}-1\right)^2\,,
\end{equation}
where $D_0(1)=2$, $D_0(2)=2 \pi$, $D_0(3)=4\pi$ and in the 1D case one must use the variables $\rho_a\rightarrow \bar{\rho}_a$ and ${g}_{\alpha\beta}\rightarrow \bar{g}_{\alpha\beta}$. Hereafter, we focus on the limit $r\ll \Delta$, where one recognizes the d-dimensional DQIM of Ref. \cite{Matthias}. The Raman coupling $\Delta$ favors a paramagnetic state where $\langle \sigma_i^z\rangle=0$ which competes with the ferromagnetic exchange $J$. The effect of dissipation will be more prominent in the 1D case, which corresponds to ohmic dissipation ($d=2,3$ rather corresponds to super-ohmic dissipation) \cite{leggett,Weiss}.

{\it Theoretical Predictions.-} For $g_{ab}=0$,  the system will be in a {\it paramagnetic} regime (since $J=0$), in accordance with Eq. (4), whereas for large $|g_{ab}|\sim g_{aa}\gg \hbar\Delta/\rho_a$, the system will be in a {\it ferromagnetic} phase. One can determine the phase (ferromagnetic or paramagnetic) of the spin system by measuring the mean occupation number $\av{n_b}=(\langle \sigma_{i}^z\rangle +1)/2$ \cite{Grangier}. The dissipation produces a damped Rabi oscillation in the paramagnetic phase \cite{Zoller2}. Below, we will assume $|g_{ab}|\ll g_{aa}$ in order to study the emergent ``dissipative'' ferro-paramagnetic transition; this results in $0<\alpha\ll 1$ (in 1D: $\alpha \approx 1/(4K)\leq 1/4$).

The nondissipative quantum Ising model is known to exhibit a second order phase transition around $|J| \sim \hbar \Delta$, separating the paramagnetic and ferromagnetic phase, and can be mapped exactly onto the classical Ising model in $d_{\text{eff}} = d+z = d+1$ dimensions; $z$ is the dynamic critical exponent that describes the relative dimensions of (imaginary) time and space. The case $z=1$ implies that time just acts as another spatial dimension. The classical model itself can be described by a $\phi^4$-theory~\cite{Subir} with well-known critical exponents~\cite{Zinn-Justin}. 

\begin{figure}[t]
  \centering
  \includegraphics[width=.65\linewidth]{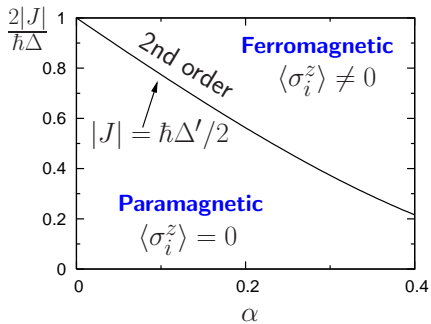}
  \caption{1D phase diagram as a function of $\alpha$ and $J$.}
  \label{fig:2}
\end{figure}

In particular, the quantum Ising chain lies in the universality class of the 2D classical model~\cite{Onsager}. With dissipation ($\alpha>0$), the critical behaviour is most profoundly changed in the ohmic case ($d=1$). The bath not only renormalizes the transverse field, $\Delta'=\Delta \left(\Delta/\omega_c\right)^{\alpha/(1 - \alpha)}$ \cite{leggett,note2}, such that the transition occurs at a smaller $|J|\sim\hbar \Delta'$ (see Fig.~\ref{fig:2}), but it also changes the {\it universality class} of the quantum phase transition. The critical exponents of this second order phase transition are independent of the value of $\alpha$ \cite{Matthias,Matthias2,Subir} and they have been derived through a dissipative $\phi^4$-theory in $d= (2 -\varepsilon)$ dimension ($\varepsilon=1$)~\cite{Georges,Matthias2} and through Monte Carlo simulations~\cite{Matthias}. Below, we summarize the main results. 

The dynamic critical exponent is equal to $z=2$ in the 1D case, whereas $z=1$ for $d=(2,3)$.  Thus, the 1D and 2D dissipative quantum Ising models should behave similarly, since they have the same effective dimension $d_{\text{eff}} = d + z \approx 3$. Experimentally, one can measure \cite{Subir,Zinn-Justin}
\begin{eqnarray}
\av{\sigma_i^z}_c &\propto& r_0^{1/\delta} \\ \nonumber
\av{\sigma^z_i} &\propto& \left|\Delta-\Delta_c\right|^{\beta} \\ \nonumber
\chi = \frac{d\av{\sigma_i^z}}{d r_0} &\propto& |\Delta-\Delta_c|^{- \gamma},
\end{eqnarray}
where $\av{\sigma_i^z}_c$ refers to the value of $\av{\sigma^z_i}=2 \av{\hat{b}^\dagger_i \hat{b}_i} - 1$ at the quantum phase transition in the presence of a finite detuning $r_0$ and $\Delta_c$ refers to the value of the Rabi frequency at the
transition. The exponents can be extracted using standard state-selective measurements of the $b$ atom mean occupation number, averaged over an homogeneous part of the lattice. By measuring the three exponents in Eq. (11), one may confirm that the phase transition is of second-order type where they are related by $\beta (\delta - 1) = \gamma$. 

In the case of the 1D dissipative system, the value of the dynamic critical exponent can be directly obtained from the equality $z=(\delta-1)/2 \sim 2$. In addition, from scaling laws, one finds 
that $\gamma=z\nu\approx 1.276$ \cite{Matthias, Matthias2, Georges} and $\beta=\nu/2\approx 0.319$, which implies that the correlation length exponent $\nu\approx 0.638$ can be inferred as well. Note that $\langle \sigma_i^z\rangle$ goes continuously to zero at a second-order phase transition which is in striking contrast to the case of a single two-level system embedded in a 1D BEC \cite{Zoller2} where a Kosterlitz-Thouless type transition takes place, and where $\langle n_b\rangle$ shows a finite jump as the detuning is changed from a negative to a positive value \cite{Karyn}. Table~\ref{tab:1} summarizes the critical exponents for various systems. 

\begin{table}[t]
  \centering
  \begin{tabular}{r|c|c|c|c|c|c|c|}
  & $\beta$ & $\gamma$ &$\delta$ & $\nu$ & $z$ \\
  \hline
  Dissipative $d=1$ & $0.319$ & $1.276$ & $5$ & $0.638$ & $\sim 2$ \\ \hline
  Non-dissipative $d=1$ & $1/8$ & $7/4$ & $15$ & $1$ & $1$ \\ \hline
  (Non)-dissipative $d=2$ & $0.325$ & $1.241$ & $4.82$ & $0.630 $ & $1$ \\
  \hline
    \end{tabular}
\caption{Critical exponents of quantum Ising models.} 
\label{tab:1}
\end{table}

In short, we have shown that a spin-boson mixture of cold atoms can be used to engineer the quantum Ising model in a dissipative bath. System parameters can be tuned to reach the quantum critical regime and critical exponents can be measured. Finally, the spin dynamics of the DQIM remains an open and challenging question.

Discussions with P. Zoller are acknowledged. This work was supported by NSF through the Yale Center
for Quantum Information Physics.


\begin{thebibliography}{10}
\bibitem{note}
The main dissipative contribution is for $i=j$ \cite{Matthias}.
\end{thebibliography}
\end{document}